\documentclass[12pt]{article}

\usepackage{bm}
\usepackage[dvips]{color,graphicx}
\usepackage{amssymb,latexsym}
\usepackage{amsmath}
\usepackage[T1,T2A]{fontenc}
\usepackage[cp1251]{inputenc}
\usepackage[english,russian]{babel}
\date{}

\begin{document}

\title{The functional mechanics: evolution of the moments of
distribution function  and the Poincare recurrence theorem.}

\author{A.\,I.~Mikhaylov. \thanks{"--- researcher,
VNIRO,laboratory of system analysis ; junior researcher, Research
and Education Center?  e-mail: mikhailov1984@gmail.com} \\
\textit{ Russian Federal Research Institute of Fisheries and
Oceanography,}\\ \textit{ 17, V. Krasnoselskaya Str., Moscow,
107140, Russia;}
\\ \textit{ Research and Education Center of Steklov
Mathematical Institute,}\\ \textit{ RAS, Moscow, 119991, Russia. } }

\maketitle

\begin{abstract}
\English One of modern approaches to a problem of the
correspondence between classical mechanics and the statistical
physics — the functional mechanics is considered. Deviations from
classical trajectories are calculated and evolution of the moments
of distribution function is constructed. The relation between the
received results and absence of paradox of Poincare-Zermelo in the
functional mechanics is discussed. Destruction of periodicity of
movement in the functional mechanics is shown and decrement of
attenuation for classical invariants of movement on a trajectory
of functional
mechanical averages is calculated.

\end{abstract}

\section{Introduction. }
\English The irreversibility problem, i.e. the problem
correspondence between irreversible in time the equations of
macroscopic dynamics (transport equations and etc.) and the
fundamental reversible microscopic equations ( Hamilton and
Schrodinger equations) is one of the most profound problems of the
physical theory. Since paradoxes of Loschmidt and Poincare-Zermelo
finding contradiction between the picture of relaxation to
thermodynamic balance and the fundamental properties of mechanical
motion were formulated, became clear that the reductionism in the
narrow sense, ie to deduce macroscopic equations only from the
elements of mechanics, rather likely, impossible, and
methodologically unacceptable. One of the manifestations of the
above-mentioned controversy is the fact that the Poincare's
recurrence theorem implies the impossibility of entropy as a
function of canonical variables, monotone increasing on the
trajectories of the system. On the other hand, the validity of
both  mechanics equations and the equations of physical kinetics
makes it necessary to construct procedure, to deduce of the
macroscopic equations from some extra prescription, and within
these prescription withdrawal must be unambiguous, and
 prescription themselves should be independent of the specific form of the
Hamiltonian system. To date,  approaches to solve the problem of
irreversibility have been proposed, and, nevertheless, the problem
is still not investigated to the end and requires additional
study. Most modern approaches consider the properties of
macroscopic dynamics of the attributes of the phase flow mechanics
system as a whole, rather than a single isolated trajectory,
substantially different in the implementation of this
methodological principle.
 The most well-known approach of this
kind is the method of Bogoluybov \cite{Bogolub}, being
constructing the  hierarchy Bogoluybov- Born - Green - Kirkwood -
Yvon (BBGKY) by averaging the multiparticle Liouville equation.
 Another original method in statistical mechanics was developed by Prigogine \cite{Prig}. This method also
 comes from the Liouville equation, but it offers quite a different
 interpretation based on an analogy between the Liouville equation and
 Schrodinger equation. Poisson bracket with Hamiltonian (Liouville operator) is considered
 as an analogue of the Hamiltonian acting in the space of states -
 distribution functions, as well as macroscopic quantities (primarily temperature and
 entropy) are the operators which do not commute with the
 Liouville operator.\Russian
  Recently, I.V.~Volovich has offered another approach to classical mechanics the so called functional mechanics\cite{Volovich_Irreversibility}.
  Не suggested way to interpret the Liouville equation as the fundamental equations of mechanics \cite{Volovich_Irreversibility1}, the distribution function is also  considered as the system state, but the state is not of the macroscopic but microscopic system.
 In the functional mechanics one considers the mean value of the coordinate when includes corrections to the Newton trajectory.
  Since the functional mechanics modifies the
  equations of classical microscopic dynamics, it is necessary to understand, to
what  observed effects  such a modification can lead. In the first
two articles (see
\cite{Volovich_Irreversibility}\cite{Volovich_Trushechkin}),
devoted to the functional mechanics, it was assumed that the
existence of deviations of functional mechanics mean from
Newtonian trajectories and within the perturbation theory  was
shown by the example of the simplest nonlinear mechanical systems
that such deviations occur. \\ \English
 In this paper
 the general formula for the corrections to
Newtonian trajectories in the form expansion in the moments of the
initial distribution and the calculation of the evolution of all
moments of higher order will be displayed. The results will be
used for proof of the destruction of the periodicity of the
functional mechanics and of non-conservation classical invariants
of  movement in the functional mechanic trajectories. This paper,
I hope, will elaborate further observable consequences of the
functional mechanics.
\section{Liouville equation and the evolution of the distribution function.}
The Liouville equation  determines the dynamics of the system  in
functional formulation of classical mechanics.

\begin{eqnarray}
{\partial}_{t} \rho + {\partial}_{i}( \rho {v}^i )=0
\label{liuv}\\
\rho(0,x)=\rho_{0}(x)\label{liuv_init}
\end{eqnarray}
 Here, $ x $ belongs to the phase space, ie is a pair
  variables $ x = (p, q) $, where $ p, q \in \mathbb {R}^{n} $, and $v^i ,( i=1,\ldots 2n ) $ is a vector field of corresponding
  Hamiltonian system with Hamiltonian $ H (p, q) $.
 \begin{eqnarray}
 \mathbf{v} = \sum_{\mu=1}^{n}(\frac{\partial H}{\partial p_\mu}{\partial}_{q_\mu}-\frac{\partial H}{\partial
 q_\mu}{\partial}_{p_\mu})\label{VF}
 \end{eqnarray}
 State of the mechanical system is
 determined by the probability density $ \rho (t,x) $.It makes sense
  to draw parallels with quantum mechanics - probability density
 is an analogue of the density matrix and the linear functionals over
 state space are meaningful observables. If in the case of quantum mechanics, observables are determined self-adjoint operators and the average value of the observable $ \langle \hat{O} \rangle = Tr \hat {O} \hat {\rho} $
 , in the case of the functional mechanics  the observables are
 c-integrable numerical\footnote{It would be natural to assume that the observables  and the state are square integrable, but in the original paper on the functional mechanics formulation of the amendments to the Newtonian trajectory is designed to enhance the state space to a space of generalized functions, which in turn  would narrow the class of observables. Although in our calculation of the corrections we shall not resort to the apparatus of generalized functions, it should be noted that the question of the class of state functions needs further clarification. } over measure $ \rho $
 functions and the average value of the observable $ \langle O \rangle = \int O (x)\rho (x) dx $.
 Since the functional mechanics  observables
 commute with each other, the problem of simultaneous measurement
 is absent--- in particular, the momentum and coordinate can be measured
 simultaneously with any desired (but not absolute, that is essential ) precision within defined properties of the instrument\footnote{Note that the quantum mechanical analogue of the axiom of measurement in the functional mechanics had not yet formulated, we can only say that, as in quantum mechanics, the measurement is a reduction of the density function (see \cite{Trush}), but the specific form of the density after a measurement and its relation to the properties of the device requires additional discussion },
  that is why the functional mechanics is classical.

  The solution of the Cauchy problem \eqref{liuv} - \eqref{liuv_init} for the Liouville equation can be written
  as
\begin{eqnarray}
\rho (t, x) = {\rho}_{0} (u(-t,x))\label{liuv_solve}
\end{eqnarray}
where $ u (-t, x) $-phase flow along the solutions of the
equations of the characteristics  , i.e. the family  of Cauchy
problem for equations of a dynamical system with all possible
initial data
\begin{eqnarray}
\dot{{u}}(t,x)={v}(u)\label{Cosh}\\
 {u}(0,x)= x \label{Cosh_init}
\end{eqnarray}

We will consider not only the Hamiltonian system, but also a wider
class of dynamical systems which preserves the phase volume
  flow of $ \det (\frac {\partial {u}^{i}} {\partial x^j}) = 1 $ for which  functional reformulation
dynamics is reasonable due to the equality
 \begin{eqnarray}
\int f(x){\rho}_{0} ({u}(-t,x))dx =\int
f({u}(t,x)){\rho}_{0} (x)dx \label{flow_cons}
\end{eqnarray}

Here $f$ is a function on the phase space. Formally, the solution
of the problem \eqref {Cosh} - \eqref {Cosh_init} can be
represented by an exponential function of a vector field

\begin{eqnarray}
{u}(t,x)=e^{t \mathbf{v}}x\label{EVF1}
\end{eqnarray}
where the exponent is the sum of series
\begin{eqnarray}
e^{t \mathbf{v}} \equiv 1 +  t v^i {\partial }_{i} +\frac{1}{2}
t^2 v^k {\partial }_{k}v^i {\partial }_{i} + \ldots
=\sum_{n=0}^{\infty}\frac{t^n}{n!} (v^i {\partial
}_{i})^{n}\label{EVF2}
\end{eqnarray}
 The exact solution is not necessarily obtained by summing the series, it
  can be obtained in any other way - by  the summation
  perturbation series, or explicitly in some
  special cases - all subsequent arguments we will
  build on the assumption that the exact solution is known.

As it is known from the theory of ordinary differential equations
function $ u (t, x) $ is continuously differentiable as many times
as $ v $, in particular, if $ v $ is analytic and that $ u (t, x)
$ analytic (see eg \cite{Tih} to <<local>> theorem and
\cite{Pontryagin} for <<global>>) on some time interval, so we can
expand this solution in a number of Taylor on the coordinate $ \xi
= x - x_0 $ in the vicinity of some point $ x_0 = \int x \rho_0
(x) dx $, chosen so that it coincided with the expectation of the
coordinates on the initial distribution.
\begin{eqnarray}
u(t,x_0 + \xi) =\sum_{\alpha \in \mathbb{N}_{0}^{D}}\frac{1}{
\alpha !}
{\xi}^{\alpha}{\partial}_{\alpha}u(t,x_0)\equiv\sum_{\alpha
}{\xi}^{\alpha}u_{\alpha}\label{Tey}
\end{eqnarray}
We have used the compact notation of the Taylor series through
multi-index $\alpha \in \mathbb{N}_{0}^{D} $ - vector non-negative
integer components, for which in addition to operations of linear
algebra operations are defined the modulus $ | \alpha | = \sum_{i=
1}^{2n} \alpha_i $, componentwise multiplication $ \alpha \beta =
\{\alpha_i \beta_i \} $, the factorial of $ \alpha! =
\prod_{i=1}^{2n}{\alpha_i}! $, as well as the erection of a vector
in the degree $ x^{\alpha} = \prod_{i=1}^{2n} x_{i}^{\alpha_i} $,
and the taking of partial derivative $ {\partial}^{\alpha} =
\prod_{i = 1}^{2n} \partial_{i}^{\alpha_i} $ order $ \alpha $.
Taylor series coefficients, ie the partial derivative order $
\alpha $, with appropriate weight, we have denoted as $ u_{\alpha}
= u_{\alpha} (t) = \partial^{\alpha} u / \alpha! $.

Now, averaging \eqref{Tey} on the initial distribution, it is easy to
calculate the average values of coordinates \eqref{coord} and
deviations of averages of Newtonian trajectories \eqref{dcoord}
through centered moments of the initial distribution of $
{M}^{\alpha} $ defined by \eqref{moment_def}, where $\langle
u(t,x)\rangle =\int u(t,x)\rho_0(x)dx$
\begin{gather}
\langle {u}(t,x)\rangle =\sum_{\alpha
}{M}^{\alpha}u_{\alpha}\label{coord}\\
\langle {u}(t,x)\rangle -{u}(t,x_0)=\sum_{|\alpha|>1
}{M}^{\alpha}u_{\alpha}\label{dcoord}\\
{M}^{\alpha}=\langle \prod_{i=1}^{D}(x^{i}-x_{0}
^{i})^{\alpha_i}\rangle =\int\prod_{i=1}^{D}(x^{i}-x_{0}
^{i})^{\alpha_i}\rho_0(x)dx
=\int\prod_{i=1}^{D}(\xi^{i})^{\alpha_i}\rho_0(x_0+\xi)d\xi\label{moment_def}
\end{gather}
Thus, the functional mechanical corrections to the classical
Newtonian trajectories are given a countable set of time functions
$ U_{\alpha} $ - all the partial trajectory classical motion of
the initial data, which were calculated on the  classical
trajectory itself. Therefore, to calculate the deviations from well-known
classical trajectory it is not necessary to find solution for all
initial data - enough to know the behavior trajectories in the
vicinity of the know one, or build a chain of variational equations
(see \cite{Pontryagin}), which defines the evolution Partial
solutions of differential equations parameters - in this case the
initial data. After desired set of functions is found, it remains
only to sum its pre-defined and time-varying weights - moments of
the initial distribution.
 We see that the trajectory of the average values of coordinates is defined by
 all moments of the distribution function at the initial moment of
 time, therefore there is no differential equation
 with a finite number of parameters that would be effect the trajectory
 averages, since the solution of this equation is uniquely determined
 by initial conditions and parameters, while the trajectory
 mean values depend on the moments of the initial distribution,
 which can be chosen rather arbitrarily. Then naturally arises the problem of tracing the evolution
 centered moments of all orders. This requires in
 the original definition of \eqref{moment_def} to make the substitution $ x \mapsto {u} (t, x) $ and input a series of \eqref{coord} into
 a new
 definition \footnote {$ M^{\alpha} (0) = M^{\alpha} $ due to the fact that $ u (0, x) = x $} of a centered moment of order $ \alpha $, and then to rearrange the multiplication and summation.
  We carry out this
 tedious calculations in three stages. At first, we transform the definition of
 centered moments with the help of \eqref{moment0}, making a
 the original definition of \eqref{moment_def} replacement $x\mapsto{u}(t,x)$
\begin{multline}
M^{\alpha}(t)=\langle ( {u}(t,x) - \langle {u}(t,x)\rangle
)^{\alpha}\rangle =\langle\sum_{\beta \leqslant \alpha}(
(-1)^{|\beta|}{u}^{\alpha -\beta}(t,x)  \langle {u}(t,x)\rangle
)^{\beta}\rangle =\\=\sum_{\beta \leqslant \alpha}
(-1)^{|\beta|}\langle {u}^{\alpha -\beta}(t,x)\rangle { \langle
{u}(t,x)\rangle }^{\beta}\label{moment0}
\end{multline}
Then we calculate the average degree of any monomial coordinates
at a certain time by
\begin{eqnarray}
\langle {u}^{\alpha -\beta}(t,x)\rangle =\langle ({\sum_{\gamma
}{\xi}^{\gamma}u_{\gamma}})^{\alpha -\beta}\rangle =\sum_{\gamma
}\langle{\xi}^{(\alpha -\beta)\gamma}\rangle u^{\alpha
-\beta}_{\gamma} =\sum_{\gamma }M^{(\alpha -\beta)\gamma}
u^{\alpha -\beta}_{\gamma}\label{monom}
\end{eqnarray}
Here the lower multi-index is  the order of the derivative, and the upper is
the power of its components.
  Finally it remains to convert the power of mean
  coordinates by selecting the coefficients of the products of the moments, for which we use the computation

\begin{eqnarray}
\langle {u}(t,x)\rangle ^{\beta} =\prod_{i=1}^{D} (\sum_{\delta
}{M}^{\delta}u^{i}_{\delta})^{\beta_i} = \sum_{\substack{\delta_k ;\\
\delta_k \leqslant\delta_{k+1} \forall k }} ((\sum _{\substack{\text{ permutations}\\1\leqslant i_k\leqslant D}} \prod_{\substack{k=1 ;\\
\delta_k \leqslant\delta_{k+1}
\forall k }}^{|\beta|} u^{i_k}_{\delta_k})\prod_{\substack{k=1 ;\\
\delta_k \leqslant\delta_{k+1} \forall k }}^{|\beta|}
{M}^{\delta_k})\label{monom_average}
\end{eqnarray}
In \eqref{monom_average} summation in the coefficient of ordered
product of the moment is made in such a way as to take all possible
permutations of an integer index $ i $ so
  that the index $ i $-th component appeared no more than $ \beta_i $ times.

Now we can construct an algorithm for  to compute every moment of
the distribution, given the initial moments.

\begin{eqnarray}
M^{\alpha}(t)=\sum_{\substack{\delta_k ;\\
\delta_k \leqslant\delta_{k+1} \forall k }}\sum_{\beta \leqslant
\alpha}(-1)^{|\beta|}((u^{\alpha -\beta}_{\gamma} (\sum _{\substack{\text{ permutations}\\1\leqslant i_k\leqslant D}} \prod_{\substack{k=1 ;\\
\delta_k \leqslant\delta_{k+1}
\forall k }}^{|\beta|} u^{i_k}_{\delta_k}))(M^{(\alpha -\beta)\gamma}\prod_{\substack{k=1 ;\\
\delta_k \leqslant\delta_{k+1} \forall k }}^{|\beta|}
{M}^{\delta_k})\label{moment}
\end{eqnarray}

Note that if the distribution has the form $\rho (x) =
\frac{1}{\varepsilon^D} \phi (\frac{x}{\varepsilon}) $, then for
the moments the following estimate on the order of $ {M}^{\alpha} \sim
\varepsilon^{| \alpha |} $ is valid, so using \eqref{moment} in the
numerical calculation, we can retain only a finite number of
terms, while maintaining an acceptable accuracy in studying the
dynamics of average values of canonical variables on time scales
comparable to the period.

As an example, we calculate the matrix of second moments of the
distribution, given the initial moments $B^{ij}$
\begin{multline}
B^{ij}(t)=\langle ( {u}^{i}(t,x) - \langle {u}^{i}(t,x)\rangle )(
{u}^{j}(t,x) - \langle {u}^{j}(t,x)\rangle )\rangle
=\\=\langle{u}^{i}(t,x) {u}^{j}(t,x)\rangle -\langle
{u}^{j}(t,x)\rangle \langle {u}^{j}(t,x)\rangle =\sum_{\alpha
,\beta}{u}^{j}_{\alpha }{u}^{j}_{\beta}(M^{\alpha +\beta}
-M^{\alpha }M^{\beta})\label{cov}
\end{multline}
The first nonvanishing term in \eqref{cov} coincides with the
expression  for convertion of a second-rank tensor under a
  coordinates change $x\rightarrow u(x)$.
\begin{eqnarray}
B_{(0)}^{ij}(t)=\sum_{2\leqslant |\alpha| +|\beta|;|\alpha|, |\beta| \leqslant 1}{u}^{j}_{\alpha
}{u}^{j}_{\beta}M^{\alpha +\beta} = \frac{\partial u^i}{\partial
x^k}\frac{\partial u^j}{\partial x^l}B^{k l}\label{cov0}
\end{eqnarray}
 Next correction  by $ \varepsilon $  to \eqref{cov0} is given by \eqref{cov1} and contains moments of
  higher order.
\begin{multline}
B_{(1)}^{ij}(t)=\sum_{3\leqslant |\alpha| +|\beta|; |\alpha|,
|\beta| \leqslant 2}{u}_{\alpha }{u}_{\beta}(M^{\alpha +\beta}
-M^{\alpha }M^{\beta}) =\\= \frac{1}{2}(\frac{{\partial}^2
u^i}{\partial x^k
\partial x^m}\frac{\partial u^j}{\partial x^l} +
\frac{\partial u^i}{\partial x^k}\frac{{\partial}^2 u^j}{\partial
x^m \partial x^l})B^{k m l} + \frac{1}{4}\frac{{\partial}^2
u^i}{\partial x^k\partial x^m}\frac{{\partial}^2 u^j}{\partial
x^l\partial x^n}(B^{k m l n} -B^{k m} B^{l n})\label{cov1}
\end{multline}
For sufficiently small times, you can use the decomposition of
exponent of a vector field into a row and to get
\begin{multline}
B_{(0)}^{ij}(t)= B^{ij} +  t(\frac{\partial v^i}{\partial
x^l}B^{l j} + \frac{\partial v^j}{\partial x^l}B^{i l})+ \\
+ \frac{t^2}{2}(\frac{\partial (v^k {\partial }_{k}v^i)}{\partial
x^l}B^{l j} + \frac{\partial ( v^k {\partial }_{k}v^j)}{\partial
x^l}B^{i l})+ t^2\frac{\partial v^i}{\partial x^k}\frac{\partial
v^j}{\partial x^l}B^{k l} +o(t^2)\label{covdt}
\end{multline}
Expression \eqref{covdt} reproduces the phenomenon is similar to a
spreading wave packet in quantum mechanics. The results obtained in this
section  allow us to monitor the process of <<spreading>> over
formally arbitrarily long time - just enough to calculate the
classical trajectory starting at the point of phase space
corresponding to the expectation of state variables over the
initial distribution, calculate the partial derivatives of the
classical equations of motion for the initial data, and specify
the initial distribution of all its moments - the formula
\eqref{moment} describes the evolution of the moments by all
orders in <<spreading>> of the initial distribution when phase flow
transferring its .
 \section{Conclusion.}
 Let's discuss the applicability of the
results. Our arguments based on two premises - the analyticity
 phase flow for a sufficiently long period of time, and
 existence of all finite moments of the distribution function. <<Global>> theorem of continuous differentiability of the solution of \eqref{Cosh} from the initial data, which we have already referred to in the preceding part of the article \cite{Pontryagin}, guarantees the preservation of smoothness noncontinuable solutions to some of its neighborhood, so it suffices to have the infinite extendability solutions of \eqref{Cosh},in order to ensure that they are analytic for analytic right-hand side. However, if the motion is finite, ie trajectory that began in
some domain $ G $ does not leave it, and the vector field defining
right-hand side of \eqref{Cosh} defined over the entire $ G $ and
has no  the singular points in $ G $ , then the solution of
\eqref{Cosh} with initial conditions in $ G $ is infinitely
extendable. For Hamiltonian systems of the above conditions follow
from the compactness constant energy surface, which neighborhood
should be chosen as an area of $ G = \{x \equiv (p, q): h-\delta
<H (p, q) <h + \delta \} $. Then the analyticity of the right side
system \eqref{Cosh} of $ G $ provides an infinite
differentiability of noncontinuable solution of \eqref{Cosh} on
initial data, and analytic $ u (t, x) $ in some neighborhood $ X_0
$, and hence the expression \eqref{Tey}, \eqref{coord} and
\eqref{moment} will be applicable to an arbitrarily long time, as
comparable to the period of return, predicted by the Poincare
theorem and beyond this period as well. On the other hand, the
existence of a weak limit distribution function (see
\cite{Kozlov}) leads to that any trajectory of the average values
of the canonical variables ends in the average values of the
coordinates of the limit distribution, and hence to an infinitely
large time, when needed take into account all the terms of the
formula $ (\ref{moment}) $ does not convenient, because it allows
to outline the asymptotic behavior. Preservation of phase volume
and finite motion are
 conditions of Poincare's recurrence theorem, but the functional mechanics predicts the occurrence of deviations \eqref{de} averaged trajectories from the starting point for the period $T(x_0)$
\begin{eqnarray}
\langle \sum_{i=1}^{D}(u^i(T(x_0),x) - u^i(0,x))^2 \rangle
=\sum_{\substack{\alpha , \beta \\1\leqslant|\alpha |, |\beta
|}}{M}^{\alpha +
\beta}\sum_{i=1}^{D}(u^{i}_{\alpha}(T(x_0),x)u^{i}_{\beta}(T(x_0),x))\label{de}
\end{eqnarray}

This effect arises because
  depending on the period from the starting point\footnote {it is clear that for points in one orbit, the
  period is the
  same} --- differentiating the periodicity condition $ u (T (x), x) = u (0, x) $ along the coordinate,
\begin{eqnarray}
\partial^{\alpha}u(0,x)=\partial^{\alpha}u(T(x),x)=\prod_{i=1}^{D}(\partial_{i}+\partial_{i}T(x))^{\alpha_i}u(t,x)
\end{eqnarray}
 we get the non-periodicity of higher derivatives $ u_{\alpha} (T (x), x) \neq u_{\alpha} (0, x) $, while the global periodicity (the case $ \partial_ {i} T (x) = 0 $ for $\forall x $) in the functional mechanics is preserved.

 The destruction of the periodicity of the functional mechanics leads to nonconservation of the trajectories of the average values of coordinates of the integrals of motion of classical
  system - using arguments similar to the calculation of the evolution
  moments, we can obtain an expression \eqref{deI} for amendments to the evolution
  invariants of the classical system.

\begin{multline}
\delta I =I(\langle u(t,x)\rangle)-I(u(t,x))=\sum
(\sum_{|\alpha|\geqslant
1}u_{\alpha}M^{\alpha})^{\beta}I_{\beta}=\\=\sum_{\beta}I_{\beta} \sum_{\substack{1\leqslant\alpha_k ;\\
\alpha_k \leqslant\alpha_{k+1} \forall k }} ((\sum
_{\substack{\text{
permutations}\\1\leqslant i_k\leqslant D}} \prod_{\substack{k=1 ;\\
\alpha_k \leqslant\alpha_{k+1}
\forall k }}^{|\beta|} u^{i_k}_{\alpha_k})\prod_{\substack{k=1 ;\\
\alpha_k \leqslant\alpha_{k+1} \forall k }}^{|\beta|}
{M}^{\alpha_k})\label{deI}
\end{multline}
The first nonvanishing term in the expansion \eqref{deI} is given
by \eqref{deI0} and is proportional to the covariance matrix.
\begin{eqnarray}
\delta I^{(0)} =\frac{1}{2}B^{ij}\frac{\partial^2 u^{k}}{\partial
x_i\partial x_j}\frac{\partial I}{\partial x_k}\label{deI0}
\end{eqnarray}
 Thus  the paradox arises --- the functional mechanics
average values of the invariants of motion, for example, the
energy , are conserved, but as a function of average values of the
coordinates of the invariants and, specifically, the Hamiltonian
is not preserved. It seems  natural to assumed the decrease of the
Hamiltonian path functional mechanical means for Hamiltonian
systems with compact and convex Hamiltonian (and hence convex
surface of constant energy), due to conservation of the average
value of the Hamiltonian in the transition to the limit
distribution of $ \int H (x) \bar {\rho} (x) dx = \int H (x)
{\rho}_{0} (x) dx $ and Jensen's inequality $ H (\bar {p}, \bar
{q}) \equiv H (\bar {x}) = H (\int x \bar {\rho} (x) dx) <\int H
(x) \bar {\rho} (x) dx = \int H (x) {\rho}_{0} (x) dx $, where $
(\bar {p}, \bar {q}) \equiv \bar {x} $
--- time-averaged phase variables $ (p (t), q (t)) $, and
averaging over the weak limit $ \bar {\rho} (x) $ a probability
measure $ {\rho} (u (-t, x)) $ of both the observables and their
time average are the same because \eqref{limit}, where we have used
invariance of the weak limit of the phase flow:
\begin{eqnarray}
\begin{split}
\int\left(\lim_{T\rightarrow+\infty}\frac{1}{T}\int_{0}^{T}
u(t,x)dt
\right)\bar{\rho}(x)dx=\lim_{T\rightarrow+\infty}\frac{1}{T}\int_{0}^{T}\int
u(t,x)\bar{\rho}(x)dxdt=\\
=\lim_{T\rightarrow+\infty}\frac{1}{T}\int_{0}^{T}\int
x\bar{\rho}(u(-t,x))dxdt
=\lim_{T\rightarrow+\infty}\frac{1}{T}\int_{0}^{T}\int
x\bar{\rho}(x)dxdt=\int x\bar{\rho}(x)dx\label{limit}
\end{split}
\end{eqnarray}
 The above means that the prediction of dynamics are
dependent on what values are measured - the measurement of
invariants allows  to predict the state of the
classical system at subsequent times much more accurately. The latter can be
interpreted in two ways - one single measurement of invariant of
dynamic system is equivalent to multiple measurement of state
variables, or, equivalently, the invariants allow consistent
measurement, while measurement of state variables must be parallel
(simultaneously).

 Thus,  we have shown that the trajectories of functional
mechanical averages  are not periodic and do not maintain the
invariant of motion using  the formulas for evolution of the
moments of the distribution derived in the previous section, ie
the functional mechanics is free from the paradox of the
Poincare-Zermelo and describes effective dissipation,  what which
takes place in real-world mechanical systems.

\section{Acknowledgments.}

The author expresses his sincere gratitude to I.V. Volovich and
O.G. Smolyanov for suggesting the problem, and also thanks O.V.
Groshev, E.V. Piskovskii, A.S. Trushechkin and other members of
the REC MIAN special seminar on the problems of irreversibility
for valuable discussions. The author expresses special gratitude
to his supervisor, D.A. Vasilyev for consultation in the
translation this articles into English. This work was supported by
the Russian Foundation for Basic Research\Russian {\rm (}project
\No~09-01-12161-ophi\_m{\rm)}




\end{document}